\documentclass[aps,prl,twocolumn,showpacs,superscriptaddress,groupedaddress]{revtex4}
\usepackage{graphicx}
\usepackage{dcolumn}
\usepackage{bm}

\begin{document}

\title{Higher-order photon correlations in pulsed photonic crystal nanolasers}

\author{D.\ Elvira$^1$, X.\ Hachair$^1$, V.\ B.\ Verma$^2$,
R.\ Braive$^1$, G.\ Beaudoin$^1$, I.\ Robert-Philip$^1$, I.\ Sagnes$^1$,
B.\ Baek$^2$, S.\ W.\ Nam$^2$, E.\ A.\ Dauler$^3$
I.\ Abram$^1$, M.\ J.\ Stevens$^2$ and A.\ Beveratos$^{1,*}$
}

\affiliation{
$^{(1)}$ Laboratoire de Photonique et Nanostructures LPN-CNRS UPR20, Route de Nozay, 91460 Marcoussis, France\\
$^{(2)}$ National Institute of Standards and Technology, 325 Broadway, Boulder, CO 80305 USA \\
$^{(3)}$ Lincoln Laboratory, Massachusetts Institute of Technology, Lexington, MA 02420, USA \\
$^*$Corresponding author: alexios.beveratos@lpn.cnrs.fr }

\begin{abstract}

We report on the higher-order photon correlations of a high-$\beta$ nanolaser under pulsed excitation at room temperature. 
Using a multiplexed four-element superconducting single photon detector we measured g$^{(n)}(\vec{0})$ with $n$=2,3,4. 
All orders of correlation display partially chaotic statistics, even at four times the threshold excitation power.
We show that this departure from coherence and Poisson statistics is due to the quantum fluctuations associated with the small number of dipoles and photons involved in the lasing process.
\end{abstract}


\maketitle

The physics of nanoscale lasers is expected to be radically different from that of conventional lasers, due to their very high spontaneous emission ratio ($\beta$) into the lasing mode, approaching $\beta \rightarrow 1$. 
Because of the high $\beta$, the lasing threshold is low, the system is capable of lasing with a small number of emitting dipoles and a small number of photons in the cavity mode, and the lasing turn-on time is expected to be very short. 
However, the quantum fluctuations associated with the small numbers of dipoles and photons cause the output intensity of the laser to fluctuate strongly with chaotic statistics which can be present well above the stimulated emission threshold \cite{Hofmann2000,Druten2000} and may even suppress continuous-wave lasing \cite{Choudhury2009}. 
Understanding these quantum fluctuations, which requires studies of the higher-order photon statistics of the laser output, is thus crucial for the operation of nanolasers, particularly if nanoscale lasers are to be used in demanding applications, such as inter-chip optical interconnects or integrated clocks.

Several studies aimed at understanding the physics behind the dymics and the noise characteristics of nanolasers have been reported in recent years. 
Measurements of the second order correlation function g$^{(2)}(\tau)$, which is directly linked to the intensity noise properties, were reported some years ago \cite{Ulrich2007,Choi2007} and more recently the evolution of g$^{(2)}(t,\tau)$ was followed with picosecond resolution in semiconductor micropillar nanolasers operated at 4K in the weak \cite{Assman2009,Assman2010} or the strong \cite{Wiersig2009} coupling regimes. 
Lately, the noise properties of nanolasers operating at room temperature were also reported \cite{Hostein2010}. 
In most of these studies, the cavity volume was of the order of a cubic wavelength and the gain material consisted of a small number of semiconductor quantum dots.
In some recent work on plasmonic structures, lasers with even smaller volumes have been reported \cite{Noginov2009,Hill2009,Nezhad2010}, reducing further the number of emitting dipoles and photons in the cavity. 
Clearly, in such small lasers, the traditional laser theories based on the ``thermodynamic limit'' of large-ensemble averaging and continuum approximations for the gain medium and the electromagnetic field are no longer valid, as the discrete nature of the number of dipoles and photons must be taken into account explicitly \cite{Ashour2007}.

In this Letter we report an experimental study of the higher-order photon autocorrelation functions of a single-mode photonic crystal nanocavity laser, with quantum dots as its gain material, operating at room temperature, in the telecommunications wavelength range, and under pulsed excitation. 
In the second-, third- and fourth-orders of correlation, the output intensity fluctuations display a chaotic component and deviate from standard Poisson statistics, even at pump powers of over 4 times the threshold.
These noise properties can be accounted for by the quantum fluctuations arising from the small number of photons and dipoles, through a simple model that provides new insight into the physics of nanolasers. 

\begin{figure}[!ht]
   \begin{center}
   \begin{tabular}{c}
   \includegraphics[width=7cm]{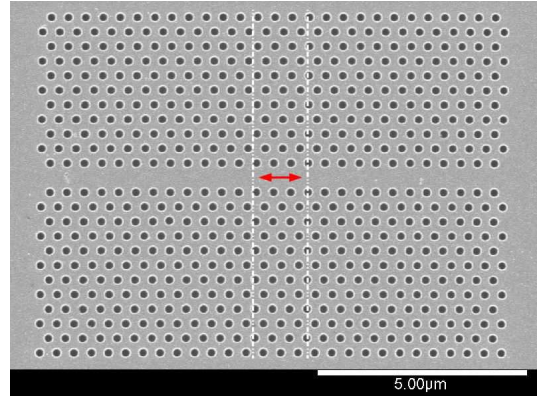}
   \end{tabular}
   \end{center}
   \caption[example]
   { \label{fig:Cavity} Double heterostructure photonic crystal cavity. White dashed lines represent the mirror position. The red arrow shows the laser cavity
     }
\end{figure}

The nanolaser cavity is formed by a photonic crystal double heterostructure \cite{Song2005} etched on a 320 nm-thick suspended InP membrane \cite{Hostein2010} and incorporating a single layer of self-assembled InAsP quantum dots \cite{Michon2008} at its vertical center plane. 
The whole structure is grown by metalo-organic chemical vapor deposition. The quantum dot density is of the order of 1.5$\times10^{9}$ cm$^{-2}$ and their spontaneous emission is centered around 1560 nm at 300 K with an inhomogeneous linewidth of about 150 nm. 
The cavity is fabricated using electron beam lithography, inductively coupled plasma etching and wet etching \cite{Talneau2008}. 
The structure consists of a W1 photonic crystal waveguide composed of one missing row of holes in the $\Gamma$-K direction of a hexagonal lattice structure with lattice constant a$_m=410$ nm, and air-hole radius of r = 0.293 a$_m$. 
The lattice constant is modified over two periods at the center of the photonic crystal waveguide along the $\Gamma$-K
direction alone, to the value of a$_l=440$ nm, thus forming a nanocavity with an effective volume of 1.3 $(\lambda/n)^3$,
enclosed by two surrounding ``mirror'' waveguides with smaller lattice constant.
The cold cavity quality factor is measured to be Q = 45000 \cite{Hostein2009}, corresponding to a cavity lifetime of $1/\Gamma_c=35$ ps.

The nanocavity laser is studied at room temperature.  
Optical excitation is provided by a pulsed Ti:Sapphire laser with a center wavelength of ~805 nm and a repetition rate of 82 MHz. 
The pump pulse is broadened to $\sim 50$ ps after passage through a 25 m length of single-mode optical fiber and is then focused to a $5 \mu$m spot on the sample with a microscope objective (NA = 0.4).  
When the quantum dots are highly excited and contain several electron-hole pairs, they emit into a broad spectrum with a lifetime of $1/\gamma_{\parallel}=400$ ps \cite{Hostein2010,Elvira2011} which feeds the cavity mode \cite{Winger2009}. 
Emission from the nanocavity laser is collected with the same objective, directed through a dichroic beamsplitter that blocks the pump light and sent to a 0.75 m grating monochromator.  
One output port of the monochromator holds an image intensified near-infrared camera, which records the emitted spectrum and power.  
Alternatively, an internal mirror can be flipped, directing light to the other output port, where it is filtered with a slit and subsequently collected into a single-mode optical fiber and sent to the detector.
The detector is a four-element superconducting nanowire single-photon detector (SNSPD) \cite{Goltsman2001} in which four independent, single-photon-sensitive elements are interleaved over a single spatial mode of the optical beam \cite{Dauler2009}.  
Each element consists of a current-biased superconducting nanowire that is driven into a resistive state, thus delivering an output voltage pulse upon absorption of one photon.  
The active area of the four-element SNSPD is matched to the mode diameter of the fiber, allowing all four interleaved nanowire elements to equally sample a single spatial mode \cite{Dauler2009,Stevens2010}.  
Fast four-channel electronics record photon arrival times on each element.  
These time-tag data are post-processed to obtain multi-start, multi-stop correlation histograms between two, three and four SNSPD elements for a continuous range of all associated time delays.  
The SNSPD is held at a temperature of $\approx$3 K in a closed-cycle helium cryocooler.  
Each element has a system detection efficiency of $\approx$ 1\% at 1550 nm and a dark count rate of  $\sim 100$ Hz. 
Details on the data processing, including the elimination of crosstalk among the interleaved detectors, are given in the annex on supplementary information \cite{SSPD}.

\begin{figure}[!ht]
   \begin{center}
   \begin{tabular}{c}
   \includegraphics[width=8.5cm]{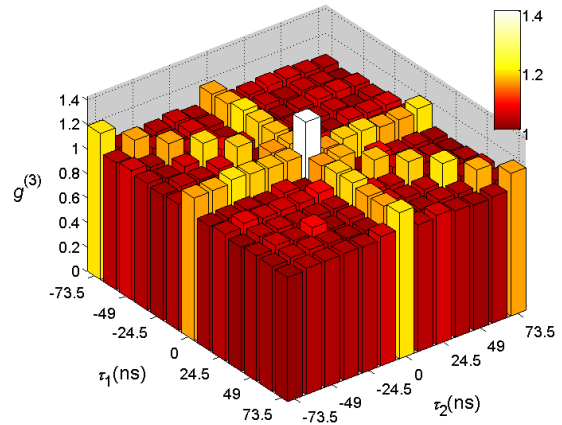}
   \end{tabular}
   \end{center}
   \caption[example]
   { \label{fig:gncurve}  Third order autocorrelation function g$^{(3)}(\tau_1,\tau_2)$ at a pump power P=2 P$_{th}$. The tallest bar is located at $\tau_1=\tau_2=0$ and represents $g^{(3)}(0,0)$. The three ridges of elevated bars that fall along $\tau_1=0$, $\tau_2=0$ and $\tau_1=\tau_2$ have values approximately equal to $g^{(2)}(\vec{0})$. The remainder of the bars have an average value of $\sim$1.
     }
\end{figure}

Figure \ref{fig:gncurve} presents a typical set of normalized experimental three-photon coincidence data, extracted from the time-tagged arrivals of photons on the four multiplexed SSPD detectors. 
It corresponds to the third order autocorrelation function, $g^{(3)}(\tau_1,\tau_2)$, in histogram form. 
The blocks at $0 \neq \tau_1 \neq \tau_2 \neq 0$ are normalized to an average value of one as the three photons detected belong to different excitation cycles; the blocks at $\tau_1=0$, at $ \tau_2 = 0$, or at $\tau_1 = \tau_2$ correspond to two photons being detected in the same excitation cycle (and the third one in another cycle) and thus give directly the value of $g^{(2)}(\vec{0})$; finally, the block at $\tau_1 = \tau_2 = 0$, where all three photons are detected in the same excitation cycle, gives the value of $g^{(3)}(0,0)$ \cite{Stevens2010}. 
For fully coherent light $g^{(n)}(\vec{0})=1$, whereas for chaotic (thermal) light $g^{(n)}(\vec{0})=n!$, for all orders $n$.
For partially coherent light, we may thus define
\begin{equation}
h^{(n)}= \frac{g^{(n)}(0)-1}{n!-1}
\end{equation}
which is a normalized ratio, indicating if the statistics are near coherent or chaotic light :
Below threshold the output light is chaotic, so that $h^{(n)}=1$, whereas above threshold, one expects a coherent laser output and $h^{(n)} \rightarrow 0$ for all $n$.

\begin{figure}[h!]
   \begin{center}
   \includegraphics[width=8.5cm]{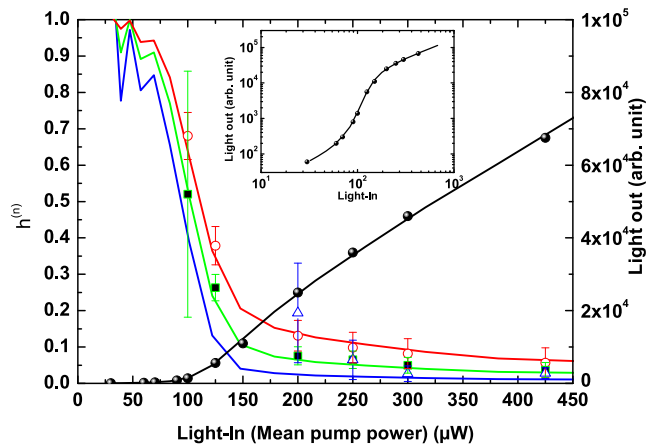}
   \end{center}
   \caption[example]
   { \label{fig:AllCurves} (Right axis) Experimental (full dots) and calculated Light-In Light-Out curve. (Left axis) Experimental values of h$^{(n)}(0)$ as function of pump power. Open circles, squares and triangles are for n=2,3,4 respectively. Full lines represent the expected h$^{(n)}(0)$ without any free parameters.}
   
\end{figure}

On Figure \ref{fig:AllCurves} we present the experimental data and the theoretical curves (obtained below) for the Light-in Light-out curve (right hand axis) and 
$h^{(n)}$ for $n =\{2,3,4\}$ (left hand axis) as a function of the pump power of the nanoscale laser under pulsed excitation.
The lasing threshold is clearly identified at the ``knee'' of the Light-In Light-Out curve (which corresponds to the inflection point of the log-log curve), at P$_{th}$=100 $\mu$W, while the $\beta$ factor is given by the ratio of the two slopes of the Light-In Light-Out curve (corresponding to the ``jump'' of that curve in log-log scale) of $\beta=0.008$.
The $h^{(n)}$, which below threshold are expected to be equal to 1, steadily decrease to reach $h^{(2)}=0.05\pm0.04$, $h^{(3)}=0.036\pm0.02$ and $h^{(4)}=0.029\pm0.015$ at $P = 4 P_{th}=400 \mu$W, all significantly higher than the value of  0 expected for coherent emission. The corresponding $g^{(n)}(\vec{0})$ are $g^{(2)}(0)=1.05\pm0.04$, $g^{(3)}(0)=1.18\pm0.10$, $g^{(4)}(0)=1.67\pm0.34$.
In other words, the intensity fluctuations still have a chaotic component at four times the threshold power, in sharp contrast with the behavior of conventional lasers (with $\beta \approx 10^{-5}$), for which all $h^{(n)}$ rapidly vanish above threshold.

The persistence of the chaotic component can be simply understood following Refs.\ \cite{Druten2000,Hofmann2000,Choudhury2009}: Because of the relatively small numbers of the emitting dipoles and photons in the cavity, a small fluctuation in these numbers introduces a non-negligible change in the stimulated emission rate, inducing chaotic (thermal) statistics even above threshold.

More quantitatively, we may model our system in terms of the traditional laser rate equations \cite{Druten2000} under the constraint that the number of excited dipoles, $N$, and the number of photons in the cavity, $s$, be integers.
\begin{eqnarray} 
\label{eq:Rate1}
\frac{dN(t)}{dt} = & P f(t) -\gamma_{\parallel}N(t) & -\beta \gamma_{\parallel} N(t) s(t) \\ 
\label{eq:Rate2}
\frac{ds(t)}{dt} = & -\Gamma_c s(t) +\beta \gamma_{\parallel} N(t) &+ \beta \gamma_{\parallel} N(t) s(t)
\end{eqnarray}
where $f(t)$ is the temporal profile of the pump pulse intensity, of integral equal to one.
For each incident power $P$, the rate equations are solved numerically by iterating over a time-step $dt$ which is small compared with the characteristic times of the system $1/\gamma_\parallel$ and $1/\Gamma_c$, and using a quantum jump approach \cite{Plenio1998} to ensure that $N$ and $s$ are integers: At each iteration step, the right hand side of Equations (\ref{eq:Rate1}) and (\ref{eq:Rate2}) is evaluated, taking into account the discrete evolution of $N$ and $s$ through a sequence of three binomial processes. First, the number of dipoles having decayed between $t$ and $t+dt$ is drawn from a binomial distribution with probability $p_1=\gamma_\parallel (1+\beta s(t))dt$ for each one of them to decay. 
Second, the decaying dipoles each produce a photon that has a probability $p_2=\frac{(1+s(t))\beta}{1+\beta s(t)}$ of entering the cavity. 
And third, the photons that have accumulated in the cavity can escape with probability $p_3=\Gamma_c dt$. 
Note that for consistency, we also take into account Poissonian statistics for the pump, although it does not alter the overall conclusions. In the end, for each incident power $P$, the g$^{(n)}(\vec{0})$ are calculated, according to 
\begin{equation}
g^{(n)}(\vec{0}) = \frac{<\int_t  \prod_{k=0}^{k=n-1} (s_i(t)-k)>}{(< \int_t s_i (t)>)^n}
\label{eq:gn}
\end{equation}
where $s_i (t)$ is the $i-th$ realization of the number of photons in the cavity, arising from the rate equations, and $<x>$ denotes the average of $x$ over all realizations.
Averages over typically 1000 realizations are performed.
As can be seen on Fig. \ref{fig:AllCurves}, the experimental data are very well described by our model for the Light-In Light-Out curve as well the second and third order correlation functions.
They are reasonably well described for the fourth order correlation function, considering the very large estimated experimental error associated with the measurements of g$^{(4)}(0)$.
Details of the calculations can be found elsewere \cite{TBP2011}.
Here, it is sufficient to point out the main differences with other methods for calculating the g$^{(n)}(\vec{0})$.
For traditional lasers with a large number of dipoles and photons in the cavity, quantum noise is introduced in the classical rate equations by means of a Langevin driving force with Gaussian statistics \cite{Hofmann2000,Druten2000}. 
This approach, however, does not take into account the discrete nature of the particles and is thus not suited for lasers with a small number of dipoles and photons. 
In addition, the usual Langevin force statistics have only a second cumulant. 
This means that the higher-order averages can only be evaluated under specific assumptions, for example, if they correspond to sums of products of lower-order averages \cite{Lemieux1999}. 
For small lasers, the discretization of the numbers of dipoles and photons is usually treated through a set of master equations for the probabilities of the states with a given number of excited dipoles $N$ and photons $s$ \cite{Choudhury2009,Ashour2007}. While this approach is ideally suited for very small systems, it requires calculating $N_0^2$ interdependent probabilities, while our method requires the computation of trajectories (realizations) with at most $N_0$ jumps. Secondly this approach allows for a more intuitive approach to understanding the underlying physical phenomena.


\begin{figure}[!ht]
   \begin{center}
   \includegraphics[width=8cm]{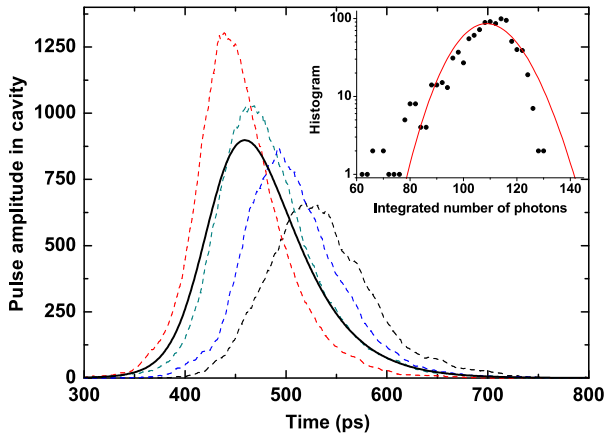}
   \end{center}
   \caption[example]
   { \label{fig:noisePtoP} Dashed lines: 4 different realizations of the stochastic rate equations at $P=1.8 P_{th}$. Full line: The mean pulse averaged over 1000 realizations. It can be clearly seen that the longer it takes for the pulse to emerge, the smaller its amplitude. (Inset) Dots: Statistical distribution of the total number of photons per pulse at $P=1.8 P_{th}$. Line: Expected distribution for Poisson statistics.}
\end{figure}

To gain a better insight into the effects of fluctuations, we present in Figure \ref{fig:noisePtoP} the evolution of the number of photons $s(t)$ for four realizations of the rate equations (dashed lines) at an excitation power $P=1.8 P_{th}$ as well as the mean of $s(t)$ over 1000 realizations (full line). 
We observe that even at pump power two times that of threshold, there are strong intensity and timing (jitter) fluctuations, which are correlated: The longer it takes for the pulse to be formed, the smaller its intensity, because in the meantime the number of excited dipoles has decreased.
The inset of Figure \ref{fig:noisePtoP} represents the statistical distribution of the integrated pulse area (in number of photons), fitted by a Poisson distribution of the same mean value. 
Clearly the number of photons in a pulse deviates from a Poisson distribution, and this will cause the autocorrelation functions g$^{(n)}(\vec{0})$ to deviate from the value of one characteristic of coherent emission. 
As can be seen in the inset, the deviation from a Poisson distribution consists of a lower probability for high-intensity events and a higher probability for low-intensity events. 
The correlation between the intensity fluctuations and the jitter implies that late starts are more probable than early starts, even though the gain medium can provide more photons at earlier times.
The physical origin of this effect may be similar to the suppression of continuous-wave lasing due to fluctuations in small lasers, discussed by Roy-Choudhoury {\it et al.} \cite{Choudhury2009}. 
Fluctuations may suppress a nascent laser pulse, leading to ``false starts" and delaying the time when the successful laser pulse takes off.
This asymmetry in the statistics of pulse intensity can be characterized by the higher (third and fourth) moments of the distribution and thus impacts the higher-order autocorrelation functions.


In conclusion we measured up the the fourth order autocorrelation function for a photonic crystal nanolaser operating at room temperature under pulsed excitation. We observe that all values of the autocorrelation function are significantly above unity even at 4 times threshold, indicating the presence of chaotic fluctuations in spite of the predominance of stimulated emission. A simple model taking into account the discrete nature of the number of photons and dipoles succesfully describes the experimental data, providing a novel understanding of the operation of the nanolasers. Measurement of the higher-order photon correlation functions thus opens the way to the study of very small lasers, whose operation deviates strongly from the ``thermodynamic limit'' that governs conventional lasers.

The authors acknowledge financial support from the Triangle de la Physique under the BIRD project and from the French National Research Agency (ANR) through the Nanoscience and Nanotechnology Program (project NATIF ANR-09-NANO-P103-36).


\begin{thebibliography}{99}


\bibitem{Hofmann2000} H.\ F.\ Hofmann, O.\ Hess, 
J.\ Opt.\ Soc.\ Am.\ B \textbf{17}, 1926 (2000).

\bibitem{Druten2000} N.\ J.\ van Druten, \textit{et al.}, 
Phys.\ Rev.\ A \textbf{62}, 053808 (2000).

\bibitem{Choudhury2009} K.\ Roy-Choudhury, S.\ Haas, and A.\ F.\ J.\ Levi, 
 Phys.\ Rev.\ Lett.\ \textbf{102}, 053902 (2009).

\bibitem{Ulrich2007} S.\ M.\ Ulrich, \textit{et al.}, 
Phys.\ Rev.\ Lett.\ \textbf{98}, 043906 (2007).

\bibitem{Choi2007} Y.-S.\ Choi, \textit{et al.}, 
Appl.\ Phys.\ Lett.\ \textbf{91}, 031108 (2007).

\bibitem{Assman2010} M.\ Assmann, \textit{et al.}, 
Phys.\ Rev.\ B \textbf{81}, 165314 (2010).

\bibitem {Assman2009} M.\ Assmann, \textit{et al.}, 
Science \textbf{325}, 297 (2009).

\bibitem{Wiersig2009} J.\ Wiersig, \textit{et al.}, 
Nature \textbf{460}, 245 (2009).

\bibitem{Hostein2010} R.\ Hostein, \textit{et al.}, 
Opt.\ Lett.\ \textbf{35}, 1154 (2010).

\bibitem{Noginov2009} M.A.\ Noginov, \textit{et al.}, Nature \textbf{460}, 1110 (2009).

\bibitem{Hill2009} M.T.\ Hill, \textit{et al.}, Opt.\ Exp.\ \textbf{17}, 11107 (2009).

\bibitem{Nezhad2010} M.P.\ Nezhad, \textit{et al.}, Nature Photonics \textbf{4}, 395 (2010).

\bibitem{Ashour2007} H. S. Ashour, M. Sokol, L. M. Pedrotti and P. R. Rice, J. Opt. Soc. Am. B, \textbf{24}, 1995 (2007)

\bibitem{Song2005} B.-S.\ Song, \textit{et al.}, Nature Materials \textbf{4}, 207 (2005).

\bibitem{Michon2008} A.\ Michon, \textit{et al.}, 
J.\ Appl.\ Phys.\ \textbf{104}, 043504 (2008).

\bibitem{Talneau2008} A.\ Talneau, K.H.\ Lee, S.\ Guilet, I.\ Sagnes, 
Appl.\ Phys.\ Lett.\ \textbf{92}, 061105 (2008).

\bibitem{Hostein2009} R.\ Hostein, \textit{et al.}, 
Appl.\ Phys.\ Lett.\ \textbf{94}, 123101 (2009).

\bibitem{Elvira2011} D.\ Elvira \textit{et al.}, to be published.

\bibitem{Winger2009} M.\ Winger, \textit{et al.}, 
Phys.\ Rev.\ Lett.\ \textbf{103}, 207403 (2009).

\bibitem{Goltsman2001} G.\ N.\ Gol'tsman, \textit{et al.}, 
Appl.\ Phys.\ Lett.\ \textbf{79}, 705 (2001).

\bibitem{Dauler2009} E.\ A.\ Dauler, \textit{et al.}, 
J.\ Mod.\ Optics \textbf{56}, 364 (2009).

\bibitem{Stevens2010} M.\ J.\ Stevens, \textit{et al.}, 
Opt.\ Expr.\ \textbf{18}, 1430 (2010). 

\bibitem{SSPD} See supplementary material for a detailed analysis.

\bibitem{Plenio1998} M.\ B.\ Plenio and P.\ L.\ Knight, Rev.\ Mod.\ Phys. \textbf{70}, 101  (1998).

\bibitem{TBP2011} A.\ Beveratos \textit{et al.}, to be submitted.

\bibitem{Lemieux1999} P.A.\ Lemieux and D.\ J.\ Durian, 
J.\ Opt.\ Soc.\ Am.\ A \textbf{16}, 1651 (1999).


\end{thebibliography}
\end{document}